\documentclass[twocolumn]{aastex701}

\begin{document}

\title{Pleiades Binary Fraction Revisited}

\author[orcid=0000-0001-5038-0089]{Dmitry Chulkov}
\affiliation{Institute of Astronomy of the Russian Academy of Sciences (INASAN)
119017, Pyatnitskaya st., 48, Moscow, Russia}

\email[show]{chulkovd@gmail.com}

\begin{abstract}

One of the nearest and best studied open clusters, Pleiades is an important cornerstone of stellar astrophysics. Despite its role as reference coeval stellar population, its multiplicity properties remain vaguely determined. The combined use of Gaia DR3 multiband photometry, astrometric parameter RUWE, non-single star solutions along with available ground-based spectroscopic, high angular resolution, and polarimetric observations enable more robust constraints on the binary star population in the cluster. Several conclusions may have broader implications for other stellar populations. Twin binaries, with mass ratio close to $q\sim 1$, tend to have lower RUWE, increasing their membership selection probability, relative to $q\sim 0.5$ systems that are disfavored. The frequently observed peak in mass ratio distribution for $q\sim 1$ binaries may be partially attributed to this bias. Photometrically fitted mass ratio is underestimated for double-lined spectroscopic binaries in agreement with other authors. Differential extinction photometrically mimics stellar binarity. An area of enlarged absorption is traced by increased polarization south of the Merope star and excluded from the analysis to avoid this bias. The fraction of systems with $q>0.6$ companions is measured to be $f=16.4\%^{+2.6}_{-0.6}$ for $m>0.5~M_\odot$ stars, which is larger than recent Gaia-based estimates, but compatible with the pre-Gaia values for Pleiades and the field population. Binary fraction shows no steady increase with stellar mass in the 0.5 -- 1.2 $M_\odot$ range, while mass ratio has a bimodal distribution with a minimum near $q\sim 0.7$.
 
\end{abstract}



\section{Introduction} 
Pleiades has been an important reference object for astronomers for centuries. Indeed, the description of the Pleiades appears in the influential work by Galileo Galilei, who described his first telescopic observations back in 1610 \citep{2025arXiv250312543L}. Stellar multiplicity is an inherent property of any stellar population \citep{2013ARA&A..51..269D}. Pleiades stars are frequently used as a testing ground for different aspects of stellar astrophysics and a detailed review of their multiplicity is desirable for credible data interpretation. 
 Despite dedicated efforts through the years, the constraints on Pleiades multiplicity remain significantly incomplete. 
 
Gaia space observatory \citep{2016A&A...595A...1G} has advanced our knowledge of the cosmos on different scales from the Solar System to the distant galaxies \citep{2025arXiv250910883P}  and brought a tremendous breakthrough for the census of Galactic open clusters \citep{2024NewAR..9901696C} and binary stars in the stellar neighborhood \citep{2024NewAR..9801694E}. 
The reported multiplicity fraction in open clusters varies in a surprisingly wide range \citep{2023A&A...672A..29C, 2023A&A...675A..89D}, which raises the possibility of unaccounted biases. Observations coming from diverse data sets allow for more robust constraints on stellar multiplicity \citep{2025A&A...693A.228C}, and this paper aims to reevaluate the binary star population in the Pleiades. 

In Section \ref{Sample} the sample of Pleiades stars is defined. Several multiplicity indicators are considered, starting from 
spectroscopic binarity in Section~\ref{SB}, followed by parameter RUWE in Section~\ref{ruwe}. Gaia non-single star solutions are explored in Section~\ref{nss}. In Section~\ref{cmd} binary parameters are retrieved from  color-magnitude diagram analysis, which is the principal method in this study. Differential extinction that may photometrically resemble binarity is discussed in Sections~\ref{photometric_outliers} -- \ref{excluded}. Finally, the binary fraction is constrained in Section~\ref{census}, and the key conclusions are briefly summarized in Section~\ref{summary}.

\section{Cluster members}
\label{Sample}
Cluster member selection usually relies on analysis of astrometric, photometric and spectroscopic data, preferably in their combination. All these data channels are compromised in case of binary or multiple stars, thus obstructing correct membership classification.  It became possible to trace Pleiades tidal arms for at least 200 pc or 70 degrees in the celestial sphere \citep{2025A&A...694A.258R} and possibly even further \citep{2024A&A...691A..28K}. However, these studies are largely based on the single-star solutions which make up the main source Gaia DR3 catalog \citep{2021A&A...649A...1G}. Relating a binary star with poorly-fitted astrometric solution to its parent cluster, especially at a large distance from the center is challenging, and to mitigate the arising biases auxiliary data sets should be considered in addition to Gaia.

Table 2 from \cite{2024AJ....168..156C} is adopted as an initial list of probable Pleiades members. It includes Gaia~DR3 entries with $G<15$ mag within $2^\circ$ of the cluster's center. Of these, 409 were deemed probable members, and membership was questioned for another 14. 
The often ignored sources with unreliable Gaia DR3 astrometric solutions are included thanks to ground-based proper motion data from the PPMXL catalog \citep{2010AJ....139.2440R} and radial velocities from several datasets. Spectroscopic data become scarce at $G\gtrsim 15$ mag, thus obstructing the extension toward fainter objects. For all considered sources, homogeneous high angular resolution observations were carried out \citep{2025AJ....169..145C}, enabling the identification of binary stars in the subarcsecond angular separation range ($ 0.1 \lesssim \rho \lesssim  1\arcsec$). 

The available data continue to grow, and the  Sloan Digital Sky Survey (SDSS) data release 19  \citep{2025arXiv250707093S} has brought new radial velocity estimates \citep{2025AJ....170...96M} based on a reanalysis of previously obtained spectra. This allows reconsidering the status of LV Tau ($G=14.89$ mag) whose astrometric solution and photometric data are consistent with cluster membership, but the reported radial velocities in Gaia DR3 and SDSS DR17 catalogs fall far from the cluster's average of $5.69\pm0.07$ km s$^{-1}$ \citep{2021ApJ...921..117T}. The new value, $1.5\pm2.2$ km s$^{-1}$, enables the reclassification of  LV~Tau as a probable Pleiades member, bringing the total number of considered sources to 410. The complete list of entries is provided in Table \ref{tab:main}. 

\section{Spectroscopic binaries}
\label{SB}

\begin{table*}
    \centering
    \begin{tabular}{cccccccccccc}
       Source  &Name& $G$ & RUWE & $P$ &$m_1 \sin^3 i$& Mass ratio &Ref& \multicolumn{2}{c}{Mass ratio} &$\sin i$\\
       Gaia DR3 && mag & & days &$M_\odot$&$q_{\rm sp}$&&$q_{\rm ph}$&$q^{\rm L}_{\rm ph}$&&\\

  {\scriptsize 66526127137440128}& Atlas, 27 Tau&3.62& 4.117&291&$5.04\pm0.17$&$0.72\pm0.01$&T25&$0.63^{+0.08}_{-0.07}$&0.55&0.95\\

 {\scriptsize 66507469798631808}&HD 23964A&6.81&0.974&16.7&1.53&0.55&T21&$0.40$&--&0.85\\

 {\scriptsize 66729880383767168}&V1229 Tau&6.82&1.145&2.46&$2.14\pm0.01$&0.70&T21&$0.54$&0.47&0.95\\

{\scriptsize 66715174415764736}&HD 23608&8.62&2-pam&7.76&$0.75\pm0.01$&0.93&T21&SB3&--&--\\

{\scriptsize 69819404977607168}&HD 23351&8.90&2.074&20.8&$1.31\pm0.03$&$0.62\pm0.01$&T21&$0.55^{+0.03}_{-0.03}$&0.53&$0.97\pm0.01$  \\

{\scriptsize 65090680344356992} &HD 23158 & 9.43 & 6.012 & 269& $1.30\pm0.18$& $0.32\pm0.02$ &T21& $0.34^{+0.08}_{-0.12}$&0.34&$1.00_{-0.03}$\\

 {\scriptsize 65207709613871744}&HD 23631B&9.74&3.377&767&$0.91\pm0.06$&$0.93\pm0.02$&T21&$0.72^{+0.03}_{-0.02}$&0.92&$0.91\pm0.02$\\

 {\scriptsize 65199978672758272} &HD 282975& 10.05 & 1.112&26.0&0.014&$0.98\pm0.01$&T21&$0.75^{+0.00}_{-0.01}$&0.67&0.24\\       
&&&&26.0&0.059&0.95&S24&&&0.18\\

{\scriptsize 65276703968959488}&HII 761&10.39&1.229&3.31&$0.42\pm0.01$&$0.68\pm0.01$&T21&$0.60^{+0.04}_{-0.02}$&0.54&$0.73\pm0.01$\\

{\scriptsize 69883417170175488} &HII 173 & 10.63 & 3.223  &481& $0.89\pm0.02$& $0.95\pm0.01$&T21&$0.84^{+0.06}_{-0.03}$&0.74&$0.97\pm0.01$\\

  {\scriptsize 66720946851771904}&HII 2027&10.71&2.419&48.6&&&T21&SB3&1&-- \\

{\scriptsize 68310359628169088}&V1084 Tau &10.85&5.585&757&$0.68\pm0.04$&$0.86\pm0.02$&T21&$0.89^{+0.10}_{-0.04}$&1.00&$0.91\pm0.02$\\

{\scriptsize 64928605459180416}&HII 2406&10.94&1.96&33.0&$0.55\pm0.04$&$0.54\pm0.02$&T21&$<0.3$&0.26&$0.83\pm0.02$\\

{\scriptsize 69876712724339456}&V1271 Tau&11.43&1.114&--&--&$0.69\pm0.09$&L25&$0.41^{+0.05}_{-0.00}$&0.56&--\\

{\scriptsize 64980278208557696}&V1065 Tau&11.96&1.228&--&--&$0.51\pm0.03$&L25&$0.84^{+0.06}_{-0.03}$&--&--\\

    {\scriptsize 66734720809017856}&HII 1348 &12.23&1.652&94.8&$0.31\pm0.01$&$0.78\pm0.01$&T21&$0.78^{+0.03}_{-0.01}$&0.78&$0.75\pm0.01$\\

  {\scriptsize 65800930496992512}&V888 Tau&12.52&1.091&3.11&0.59&0.99&T21&$0.96^{+0.00}_{-0.01}$&0.90&0.96\\

      {\scriptsize 66612473157921024} &V889 Tau& 12.60&1.053&8.58&$0.54\pm0.01$&$0.98\pm0.01$&T21&$0.89^{+0.01}_{-0.00}$&0.80&$0.92\pm0.01$\\

  {\scriptsize 66937859881182848}&V366 Tau&12.78&2.889&542&$0.48\pm0.04$&$0.82\pm0.04$&T21&$0.86^{+0.08}_{-0.02}$&1.00&$0.90\pm0.03$\\

    {\scriptsize 65005262035285888}&QQ Tau&14.26&1.12&2.46&0.10&$0.96\pm0.02$&F25&$>1$&--&$<0.59$\\

   {\scriptsize 66722355601033600} &V759 Tau& 14.70 & 1.642 & -- & -- &$0.95\pm0.08$&K21&$0.77^{+0.06}_{-0.04}$&0.68&--\\

    {\scriptsize 68259438496008832} &LM Tau&14.77&1.579&--&--&--&K21&$0.52^{+0.05}_{-0.02}$&0.46&-- \\

    \end{tabular}
    \caption{Double-lined spectroscopic binaries among Pleiades sample stars. The reported period ($P$) uncertainties are less than~1\% for all entries. $q_{\rm ph}$ is a photometrically fitted mass ratio (Section \ref{cmd}), $q^{\rm L}_{\rm ph}$ is a best-fit estimate from \cite{2025AJ....169..116L}.  Values of $q_{\rm ph}<0.5$ are generally unreliable; $q_{\rm ph}$ is not estimated for spectroscopic triple systems (SB3). References for spectroscopic solutions: F25 -- \cite{2025A&A...698A...7F}, K21 -- \cite{2021AJ....162..184K},  L25 -- \cite{2025ApJS..276...11L}, S24 -- \cite{2024MNRAS.534.3999S}, T21 -- \cite{2021ApJ...921..117T}, T25 -- \cite{2025ApJ...990..107T}. For HD 282975 two options are shown. Mass ratios from K21 and L25 are not based on a complete orbital solution and, hence, less reliable; they are omitted from the comparison with $q_{\rm ph}$ in Section \ref{photometric_validation} and Figure \ref{fig:q_compare}. For V1065 Tau and QQ Tau, $q_{\rm ph}$ is probably overestimated due to enlarged extinction (Section \ref{extinction}). The 
    $\sin i$ column is calculated from Equation \ref{eq:inclination}; $i=108^\circ$ is adopted for Atlas \citep{2025ApJ...990..107T}.} 
    \label{tab:SB2}
\end{table*}

The observation of periodic variations in the position of spectral lines due to Doppler shift is one of the fundamental methods to detect binarity \citep{2025arXiv250400548S}. The discovery and parametrization of spectroscopic binaries depend on many factors which impose a heavy bias for the obtained subsample \citep{2021ApJ...921..117T}, and many genuine systems are hardly detectable spectroscopically, such as ones with low inclination (face-on) orbits. Thus, rapid rotation typical for Pleiades stars causes line broadening which complicates the radial velocity measurements \citep{2020ApJ...901...91T}.

In a favorable case, the radial velocities of both components are measured from the observed spectra. 
The review of known double-lined (SB2) systems restricted to our sample is given in Table \ref{tab:SB2}. The mass ratio of components $q=m_2/m_1, m_2 \leq m_1$ may be estimated for a system with an orbital solution \citep{2020svos.conf..329S}, which is available for 18 of the sample systems. Two sources, HD 23608 and HII 2027, show flux contribution from a third star, making them SB3 systems; their outer companions are resolved \citep{2025AJ....169..145C}.

Generally, binarity can be detected in a single epoch spectroscopic observation \citep{2018MNRAS.473.5043E}. Ten sample sources were revealed as possible SB2 based on APOGEE data \citep{2021AJ....162..184K}, with two of them, V759 Tau and LM Tau, currently without orbital solution in the literature. Analysis of LAMOST medium-resolution data from \cite{2025ApJS..276...11L} brought four SB2 candidates, including already known V888 Tau and V889 Tau. For V1065 Tau and V1271~Tau, the mass ratio is estimated based on radial velocities of components $v_1$ and $v_2$, cluster mean and dispersion from \cite{2021ApJ...921..117T}, $s=5.69 \pm 0.48$~km~s$^{-1}$, with a method from \cite{1941ApJ....93...29W}: 

\begin{equation}
q=\frac{v_1-s}{s-v_2}
\end{equation}

\cite{2025ApJS..277...15J} after analysis of low resolution LAMOST spectra claimed binarity for 69 out of 410 sample objects based on training set with systems of mass ratios within $0.71 <q< 0.93$ range, 45 of them have low RUWE and no known resolved companions within $4\arcsec$ making them a convenient target for photometric confirmation (Section \ref{cmd}). However, only 16 sources show photometrically fitted $q>0.4$  meaning that the majority of sources are indistinguishable from single stars, and this list is likely dominated by the false positive contamination and is not considered further.

From an orbital solution of SB2 systems the $m_1 \sin^3 i$ value can be derived, while inclination ($i$) and component masses ($m_{1,2}$) remain unknown \citep{2020svos.conf..329S}. The availability of photometrically fitted masses ($m_{\rm ph}$) derived in Section \ref{cmd} allows a crude, model-dependent estimation of $\sin i$, which is given in Table~\ref{tab:SB2}:
\begin{equation}
    \sin i = \sqrt[3]{\frac{m_1\sin^3 i}{m_{\rm ph}}}
\label{eq:inclination}
\end{equation}

\begin{figure}[ht!]
\plotone{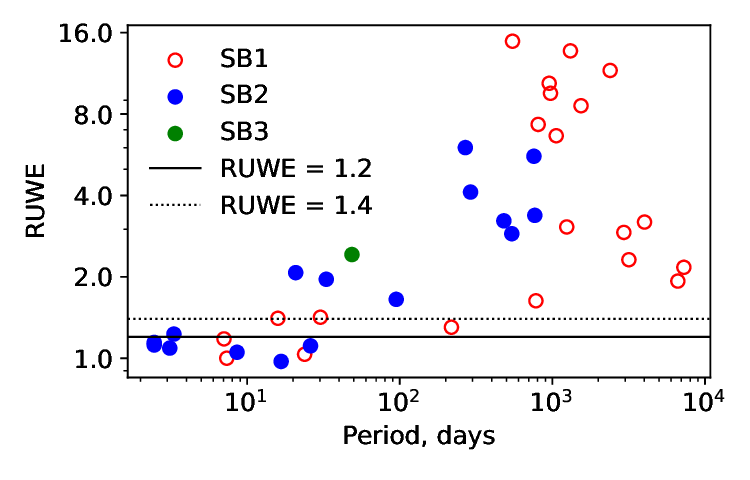}
\caption{Reported orbital periods and RUWE values for spectroscopic systems. RUWE $<1.2$ cutoff is used in Section~\ref{subsample} to select sources with reliable astrometric solutions.}  
\label{fig:SB}
\end{figure}

For single-lined systems (SB1), only the value of $m_2 \sin i$ is derived, and $q$ is not known from the orbital solution. Considering tables 6 and 12 from \cite{2021ApJ...921..117T}, there are 21 SB1 binaries in the sample. While SB2 systems have a strong preference for large mass ratio, as lines from both components should be recovered in the spectra, SB1 binaries can have much lower secondary masses. Indeed, for 11 out of 21 entries the inferred value falls into $0.05 M_\odot \lesssim m_2 \sin i \lesssim 0.2 M_\odot$ range, making their photometric confirmation very hard. However, a large $q$ remains possible, as for V1282 Tau \citep{2020ApJ...898....2T}, where the primary star has a rapid rotation. SB1 systems tend to have larger periods (Figure \ref{fig:SB}) because smaller orbital velocities reduce the chance to detect lines from both components separately.

\section{RUWE}
\label{ruwe}

\subsection{Binarity indicator}
Parameter RUWE (renormalized unit weight error) is a goodness-of-fit statistic used in Gaia catalogs that is a well-known indicator of stellar multiplicity \citep{2021ApJ...907L..33S}. Its interpretation is not straightforward, as it involves at least two different modes for resolved and unresolved sources \citep{2020MNRAS.496.1922B}. 

\begin{figure*}[ht!]
\plotone{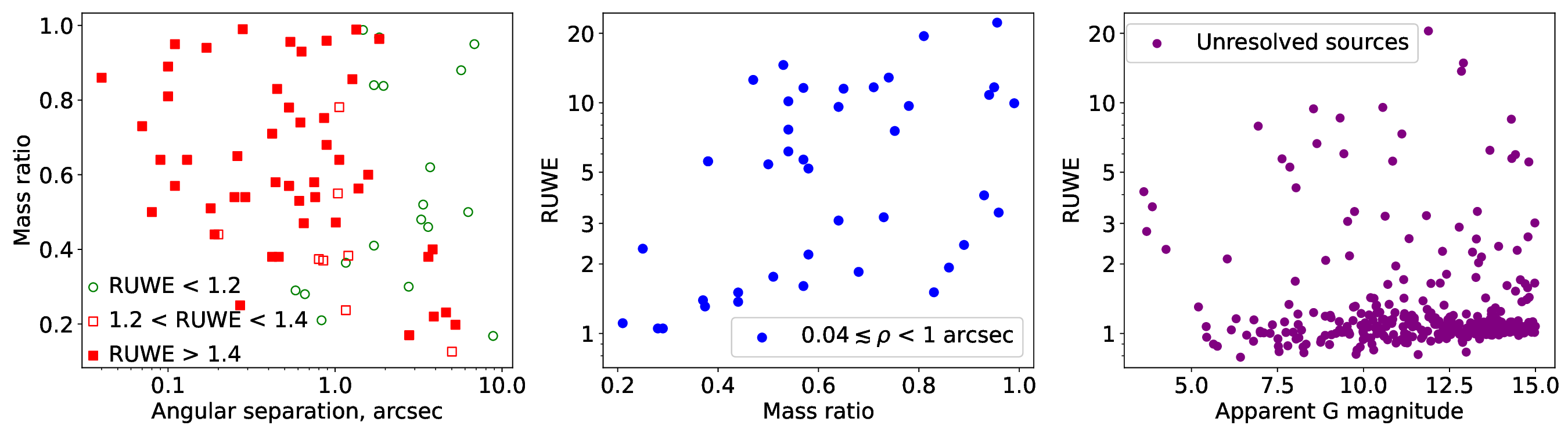}
\caption{Left: RUWE for resolved sources depending on separation and mass ratio. When both components appear in Gaia~DR3, RUWE of the brighter star is shown. Middle: RUWE for resolved sources with $\rho<1\arcsec$; observational data are from \cite{2025AJ....169..145C}. Right: RUWE for sources without any resolved companions within $4\arcsec$ depending on $G$ magnitude.}  
\label{fig:resolved}
\end{figure*}

Gaia is an optical telescope with image sampling resolution of $0.059 \times 0.177$ arcsec per pixel depending on scan direction \citep{2023A&A...674A..25H}. For marginally resolved systems the point spread function is disturbed by the companion causing a photocenter displacement, and RUWE excess is consistently observed for pairs with angular separation within one arcsec unless the secondary flux is too low (Figure~\ref{fig:resolved}). The vast majority of binaries with this mode of RUWE excess should be resolved during high resolution survey by \cite{2025AJ....169..145C}.

Another reason for RUWE increase is a wobble of the photocenter (center of light) relative to the system's center of mass \citep{2020MNRAS.495..321P}. Gaia DR3 main source catalog assumes single-star solutions \citep{2021A&A...649A...2L}, which are appropriate when the system's center of mass coincides with the photocenter observed by Gaia. This condition is not met by binary systems, which causes astrometric deviations encoded in RUWE increase. The largest shift is expected for systems with periods around Gaia DR3 observational baseline of 34 months \citep{2022MNRAS.513.2437P}, which can be noticed for spectroscopic binaries (Figure \ref{fig:SB}). RUWE excess persists both for SB1 and SB2 systems, including pairs with low secondary mass and mass ratio. A binary with equal components and $q=1$ should be perfectly fitted by single-star model and have low RUWE.

\cite{2024A&A...688A...1C} explored the threshold recommended to separate single objects from binary systems. It slightly varies across the Pleiades field, and a RUWE value of 1.2 is adopted as an upper limit for a reliable astrometric solution consistent with a single star model in this paper, which is stricter than the frequently used RUWE~$=1.4$ cutoff \citep{2021A&A...649A...5F}. 

\subsection{Forward modeling of RUWE}
\label{Unlimited}
\cite{2024A&A...688A...1C} further developed a forward model allowing RUWE prediction based on binary star parameters. Two obstacles arise to use it for credible investigation of sample objects. First, it was found that marginally small variation of input parameters often leads to large scatter of predicted RUWE values, which may be a model artifact. Another problem is the lack of reference objects for solid verification of model predictions. Even for SB2 binaries, an ambiguity remains, since their orbital orientation is not known, as longitude of ascending node $ \Omega$ and orbital inclination~$i$ remain undetermined \citep{2020svos.conf..329S}, although the latter can be roughly estimated from Equation \ref{eq:inclination}.

\begin{figure*}[ht!]
\plotone{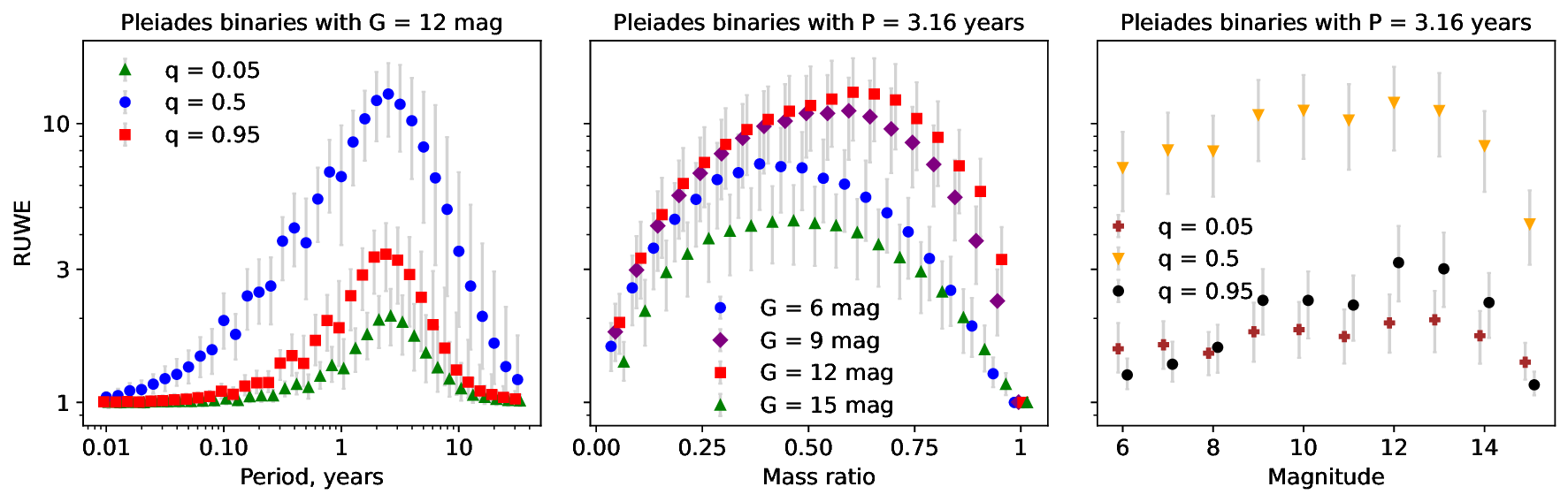}
\caption{Prediction of RUWE values for Pleiades stars with the \textit{GaiaUnlimited} package (\cite{2024A&A...688A...1C}, Section~\ref{Unlimited}). The RUWE increase for marginally resolved sources that affects binaries with $P\gtrsim 20$ years is not considered.}
\label{fig:Forward}
\end{figure*}

Simulations are carried out with \textit{GaiaUnlimited} package \citep{2024A&A...688A...1C} to study the behavior of RUWE at least qualitatively. Pleiades isochrone from \cite{2025AJ....169..116L} is used to relate stellar mass to luminosity; companions with $m_2<0.1 M_\odot$ are considered non-luminous. The parallax is set to a sample mean of $\varpi=7.37$ mas (Section \ref{subsample}), celestial coordinates of the cluster center are used: $\alpha=56.74^\circ$, $\delta=24.09^\circ$. Fixed values of apparent magnitude, mass ratio, and period are selected, eccentricity and orientation angles are distributed uniformly following the table 1 in \cite{2024A&A...688A...1C}.
A total of 1000 runs were made for each combination of parameters; the median along with 0.159 and 0.841 quantile values are shown in Figure~\ref{fig:Forward}.

\subsection{RUWE behavior}
\label{forward}

Strong dependence on binary period is predicted with a maximum around a value of 3 years due to observational time frame of Gaia DR3. While the closest systems keep a low RUWE, already at $P \sim 10$ days
binarity induces RUWE excess in many simulations. Empirically for SB1 binary HII 571 with $P=15.9$ days the RUWE value is slightly above 1.4 threshold (Figure \ref{fig:SB}). The projected decline for systems with $P \gtrsim 30$ years is largely misleading, as the model does not account for resolved binarity. Thus, V1282 Tau with an 18 year period has a $0.06\arcsec$ semimajor axis matching Gaia along-scan resolution \citep{2023A&A...674A..25H}. In practice, the resolved pairs show RUWE excess up to $2\arcsec$ separation (Figure \ref{fig:resolved}). 

There is a dependence on apparent magnitude as the sensitivity drops at the faint end of $G=15$ mag, with a broad maximum around $G\sim 12$ mag. Among unresolved sources, only three have RUWE $>10$, their magnitudes are within $11.9\lesssim G \lesssim 12.9$ range (Figure~\ref{fig:resolved}).

The probability of detecting a companion depending on its mass ratio is of special interest. RUWE is most sensitive to $q \sim 0.5$ systems. It appears that the distribution maximum is shifted toward $q>0.5$ for $G=9$ and 12~mag, while at the faint and bright ends it peaks at lower $q$ (Figure \ref{fig:Forward}). According to simulations, RUWE excess is noticeable at least up to $q\sim0.03$, and the secondary masses around $0.02~M_\odot$, which are deep into the brown dwarf range, are detectable. The longer time coverage of Gaia DR4 hopefully will increase sensitivity and allow us to probe giant exoplanets in the Pleiades. The decrease of RUWE for binary stars with $q\sim 1$ may create a bias for member selection and multiplicity statistics in different stellar populations as discussed below.

 \subsection{Impact on cluster member selection}
\label{q_bias}

The amount of photocenter wobble relative to center of mass is smaller for binaries with $q\sim 1$, so they are more likely to have astrometric solution compatible with single star model and have a low RUWE. This effect may have broader consequences for the observed mass ratio distribution in open clusters and other stellar populations accessed with Gaia. Cluster member selection depends on many factors, such as proper motion and field stellar density, and the feasibility of credible classification varies between different stellar groups. Genuine members have a better chance to be identified if they have a reliable low-RUWE solution. Binarity introduces deviations to astrometric fit which may prevent a source from being recognized as cluster member.  Objects with larger RUWE have a lower chance to be properly classified, even if no explicit cutoff is made. Thus, systems with $q\sim1$ are more likely to get a reliable astrometric solution and to be selected as cluster members, while binaries with $q\sim0.5$ are disfavored. A distinct peak at $q=1$ arising during the photometric data analysis of multiplicity \citep{2025ApJ...989..104C} may be induced by the preferential selection of such near-twin binaries. The selection bias affects clusters to varying degrees and likely contributes to the large reported spread of multiplicity fraction \citep{2023A&A...672A..29C}.

\section{Gaia DR3 NSS solutions}
\label{nss}

\begin{deluxetable}{ccccccr}
\addtolength{\tabcolsep}{-0.25em}
\tablecaption{Gaia DR3 non-single star solutions}
\tablehead{
    \colhead{Source}     & \colhead{$G$} & \colhead{\scriptsize RUWE} & \colhead{nss} & \colhead{\scriptsize SB} & \colhead{$P$} & \colhead{$q_{\rm ph}$}}
    \startdata
Gaia DR3& mag&&\multicolumn{2}{c}{Type}&days&\\
\scriptsize66507469798631808&6.81&0.974&SB2&2&16.7&0.40\\
\scriptsize66729880383767168&6.82&1.145&SB2&2&2.46&0.54\\
\scriptsize64898368889386624&8.12&10.395&acc9&1&953&$<0.4$\\
\scriptsize65275501377570944&8.55&9.413&acc9&&&$<0.4$\\
\scriptsize65090680344356992&9.43&6.012&Orb&2&269&$<0.4$\\
\scriptsize65207709613871744&9.74&3.377&acc7&2&767&0.72\\
\scriptsize66771146429454592&10.30&1.419&SB1&1&30.2&$<0.4$\\
\scriptsize65276703968959488&10.39&1.229&SB1&2&3.31&0.60\\
\scriptsize69829609819915648&10.56&9.547&tSB1&1&972&$<0.4$\\
\scriptsize69883417170175488&10.63&3.223&ASB1&2&481&0.84\\
\scriptsize65232105028172160&10.64&2.921&acc7&1&2940&$<0.4$\\
\scriptsize66863058730966528&10.71&1.632&acc9&1&780&$<0.4$\\
\scriptsize68310359628169088&10.85&5.585&acc9&2&757&0.89\\
\scriptsize69864313155605120&11.01&1.406&SB1&1&15.9&0.48\\
\scriptsize66747816167123712&11.90&1.18&SB1&1&7.05&$<0.4$\\
\scriptsize66519530066802944&12.30&2.27&acc9&&&$<0.4$\\
\scriptsize66937859881182848&12.78&2.889&acc9&2&542&0.86\\
\scriptsize65660158649542784&12.85&13.699&acc9&1&1313&0.73\\
\scriptsize66816187748666624&12.90&14.871&ASB1&1&548&$<0.4$\\
\scriptsize66944422591073024&13.31&3.381&acc9&&&$<0.4$\\
\scriptsize64841503521468544&14.30&8.486&acc9&&&0.56\\
\scriptsize65229734206151680&14.78&2.622&acc7&&&0.61\\
\enddata  
\tablecomments{Complete list of nss solutions for sample objects; the nss acronyms are defined in Section~\ref{nss}. The SB and $P$ columns refer to the SB1 or SB2 binarity type and period from \cite{2021ApJ...921..117T}, $q_{\rm ph}$ is the mass ratio estimate derived from the CMD analysis (Section~\ref{cmd}).}
    \label{tab:nss}
\end{deluxetable}

\begin{figure*}[ht!]
\plotone{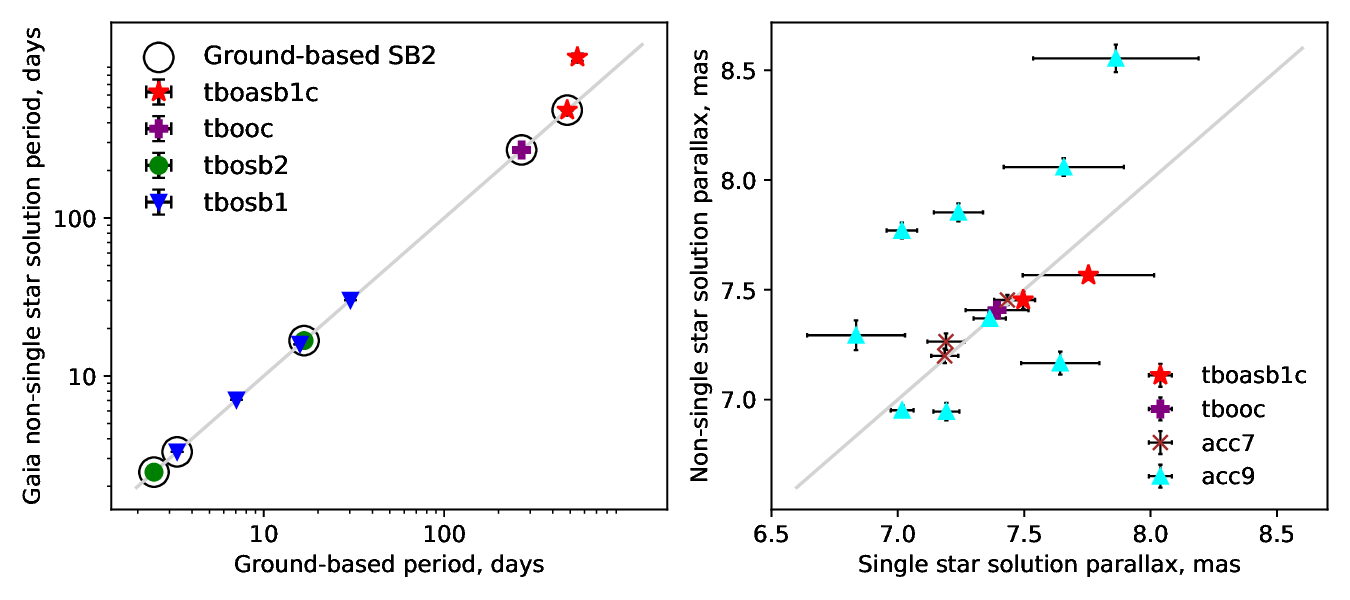}
\caption{Left: orbital periods according to Gaia DR3 nss solutions (Section \ref{nss}) and ground-based estimates from \cite{2021ApJ...921..117T}. The error bars are smaller than symbol sizes. Right: parallaxes from the main source catalog and nss solutions.}  
\label{fig:nss}
\end{figure*}

Gaia DR3 \citep{2023A&A...674A...1G} includes around $8\cdot10^5$ dedicated non-single star (nss) solutions representing a small fraction of the full catalog of $1.8\cdot10^9$ sources. Their selection  function is quite sophisticated \citep{2024OJAp....7E.100E}. Thus, face-on orbits are negatively biased \citep{2025ApJ...981L..23M}. Solutions of seven different  types \citep{2023A&A...674A...9H, 2025A&A...693A.124G}  are listed for 22 sample sources in Table~\ref{tab:nss}: 
\begin{itemize}
    \item Combined astrometric + single-lined spectroscopic orbital model (\textit{tboasb1c}, ASB1) -- 2 entries. 
    \item Orbital model for an astrometric binary with Campbell orbital elements (\textit{tbooc}, Orb)  -- 1 entry
    \item Single-lined spectroscopic binary model -- 4 entries (\textit{tbosb1}, SB1)
    \item Double-lined spectroscopic binary model -- 2 entries (\textit{tbosb2}, SB2)
    \item Spectroscopic binaries compatible with single-lined first degree trend (\textit{linspec1}, tSB1) -- 1 entry
    \item Acceleration model with 7 parameters (\textit{acceleration7}, acc7) -- 3 entries
    \item Acceleration model with 9 parameters (\textit{acceleration9}, acc9) -- 9 entries
\end{itemize}

Among 22 entries, 17 are spectroscopic binaries from ground-based observations \citep{2021ApJ...921..117T}. For the first four  types, nss solutions include orbital periods; in 8 out of 9 cases, they agree with existing estimates (Figure \ref{fig:nss}). The exception is V338 Tau ($G=12.90$ mag), an SB1 binary with $P=548\pm1$ days. Its reported period in Gaia is twice as large, $P=1042\pm55$ days, which is likely an artifact of unevenly sampled data. 

Astrometric solutions for 15 systems include parallaxes $\varpi$ that account for binarity. One could expect these values to converge better around the cluster mean of $\varpi=7.37$ mas in comparison to single star model solutions for the same sources in the main catalog, but the results contradict this expectation. Although the reported parallax uncertainties for individual entries are smaller (Figure \ref{fig:nss}), their median $\varpi$ differs more from the expected value and the standard deviation of the nss solutions is larger: $\varpi=7.41 \pm 0.43$ instead of $\varpi=7.36 \pm 0.29$~mas. The sample is small for definitive conclusions, and while the peculiar spatial distribution of these sources is possible, it seems nss solutions do not necessarily improve parallax reliability. For a larger set of triple systems, \cite{2024PASP..136i4203N} found true parallax errors are roughly ten times larger for single star solutions than for orbital nss solutions, and five times larger than for acceleration solutions. 

Four entries have masses of both components derived in Gaia DR3 \citep{2023A&A...674A..34G}. The mass ratios obtained for sources with combined astrometric and SB1 solutions (\textit{tboasb1c}) contradict the existing estimates. 
The reported mass ratio for HII 173 
in Gaia DR3, $q\sim 0.52$, is refuted by the $q \sim 0.95$ estimate for a known SB2 system (Table \ref{tab:SB2}). For SB1 binary V338~Tau, the nss solution infers $q\sim 0.86$, while $q<0.4$ is expected from the multicolor photometry (Section~\ref{cmd}). 

Inclination estimates from Equation \ref{eq:inclination} for the two SB2 systems roughly agree with nss solutions. HD 23158 has an orbital solution  with $i=109 \pm 1^\circ$ ($\sin i \sim 0.95$), while Equation \ref{eq:inclination} suggests an edge-on orbit with ${\sin i \sim 1.00}$.  For HII 173 with \textit{tboasb1c} type, Gaia provides $i=86\pm 2^\circ$ ($\sin i \sim 1.00 $) versus $\sin i= 0.97$ value from Equation~\ref{eq:inclination}.

\section{Color-magnitude diagram analysis }
\label{cmd}

\begin{figure*}[ht!]
\plotone{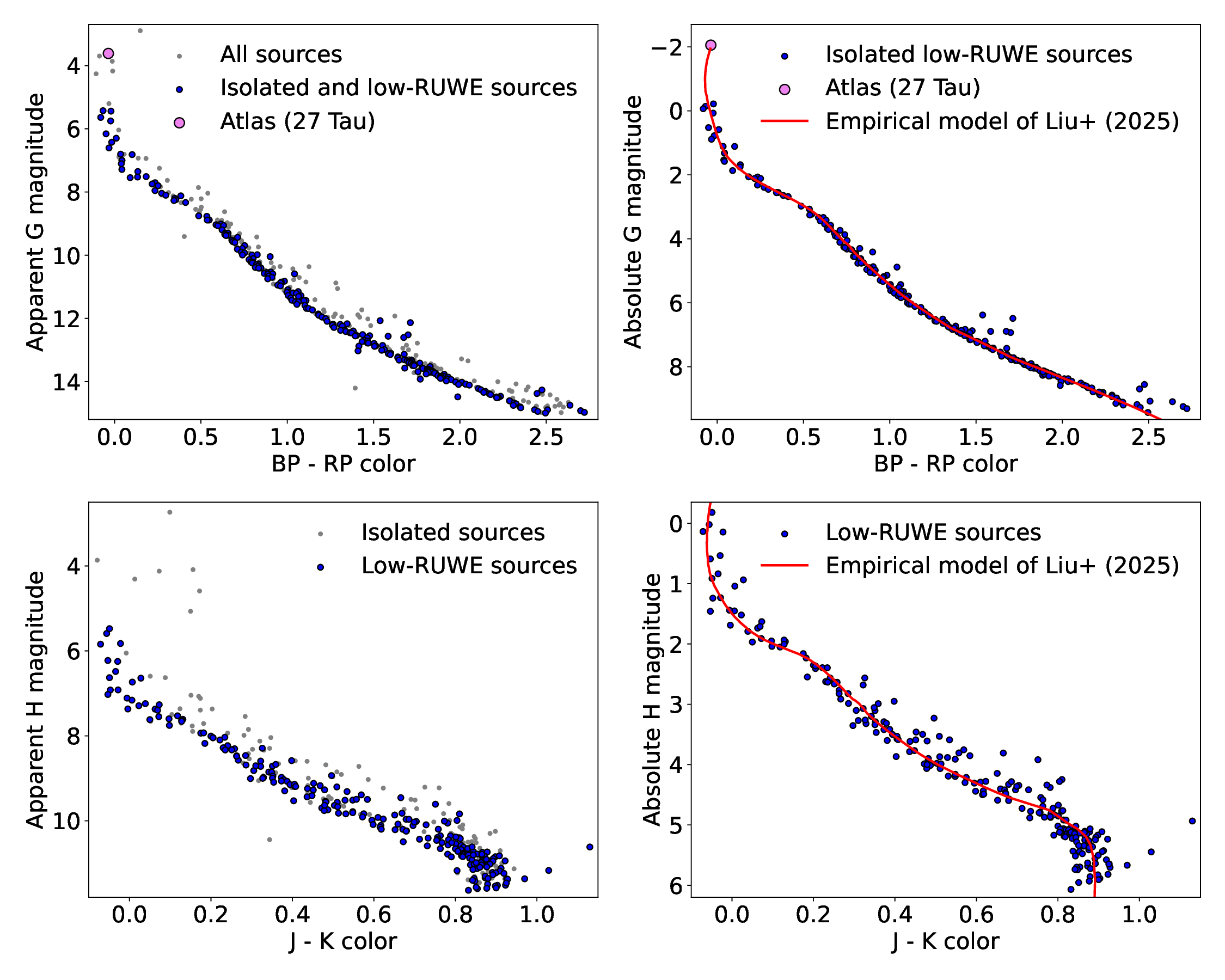}
\caption{Pleiades stars in the CMD using Gaia (top) and 2MASS (bottom) photometric data. Apparent and absolute magnitudes are shown in the left and right panels, respectively. Extinction correction is applied to the model curves when deriving absolute magnitudes. Isolated sources are defined as those with a single Gaia counterpart within $4\arcsec$. A separate threshold of RUWE $<1.2$ is adopted (Section~\ref{subsample}). For Atlas, a distance estimate of 136.2 pc is used \citep{2025ApJ...990..107T}.} 
\label{fig:CMD}
\end{figure*}

Pleiades is a typical example of a coeval simple stellar population that shares a common age, metallicity and remains relatively compact in its central part. As a result, cluster members have distinct mass-magnitude and color-magnitude relations (Figure \ref{fig:CMD}). The analysis of the color-magnitude diagram (CMD) is a long-standing way of cluster characterization,  enabling the study of its multiplicity properties, as binaries with large enough secondary flux show an offset relative to the single stars sequence \citep{1975PASP...87..707B}. High-quality Gaia photometric data in $G$, $BP$ and $RP$ passbands \citep{2021A&A...649A...3R} are widely used to put constraints on binary population within open clusters \citep{2023A&A...672A..29C}. 

The sensitivity of the photometric method depends on the component mass ratio, while being independent of the orbit's size and orientation for unresolved sources. It becomes progressively less reliable for low-mass companions as secondary component flux becomes harder to observe in front of a primary star. This section explores an application of the CMD analysis to the Pleiades. 

A subsample of sources with reliable astrometric and multicolor photometric data is defined in Section \ref{subsample} and used to adopt a single-star isochrone and estimate mass ratio in Section \ref{q_calculation}. The results are validated for known spectroscopic systems in Section \ref{photometric_validation}. The CMD analysis has its caveats and outlier sources are discussed in Section~\ref{photometric_outliers}. Differential interstellar extinction is proposed to cause the discrepancies in Section~\ref{extinction}. An area of enlarged extinction which biases the derived binary properties is outlined in Section \ref{excluded}. Finally, the revealed binary star candidates are presented in Section~\ref{unresolved}.

\subsection{Subsample with reliable Gaia data}
\label{subsample}
Besides multiplicity, several factors affect the source's position in the CMD. One is purely geometrical. The tidal radius of Pleiades ($\sim 13$ pc, \cite{2001AJ....121.2053A}) is of the order of one-tenth the distance to the Sun. This causes a sizable scatter in the CMD due to variations in individual stellar distances. Accurate parallax ($\varpi$) estimation enables tackling this problem by switching from apparent ($G$) to absolute magnitude:
\begin{equation}
\label{eq:modulus}
    G_{\rm abs}^*=G+5\log\varpi-10
\end{equation}
where $\varpi$ is measured in mas ($10^{-3}$ arcsec). Note that here $G_{\rm abs}^{*}$ is uncorrected for interstellar extinction, as absorption is instead taken into account in model isochrones. Similar conversion can be applied to arbitrary passbands. However, reliable parallaxes are not available for every star in Gaia DR3, and the appropriate subsample is defined by two separate conditions:
\begin{itemize}
\item{Sources with $\textrm{RUWE}<1.2$ are selected, as larger values may indicate the presence of a companion (Section \ref{ruwe}) which can disturb parallax estimation.}
\item{Entries with another Gaia DR3 source within $4\arcsec$~radius are excluded to avoid flux contamination from a nearby star in the $BP$ and $RP$ passbands.}
\end{itemize}
 
These criteria are met by 231 sources shown with blue dots in the CMD (Figure \ref{fig:CMD}). The median of the reported parallax uncertainties $\sigma_\varpi/\varpi$ is 0.26\%, and the precision of absolute magnitude calculation through Equation \ref{eq:modulus} is less than 0.01 mag for 94\% of entries. The median distance, estimated as $1/\varpi$, is 135.6 pc for this subsample; the 0.159 and 0.841 quantile values are 133.7 and 138.5~pc, respectively. These distances serve as references for sources with unreliable astrometric solutions.

\begin{figure*}[ht!]
\plotone{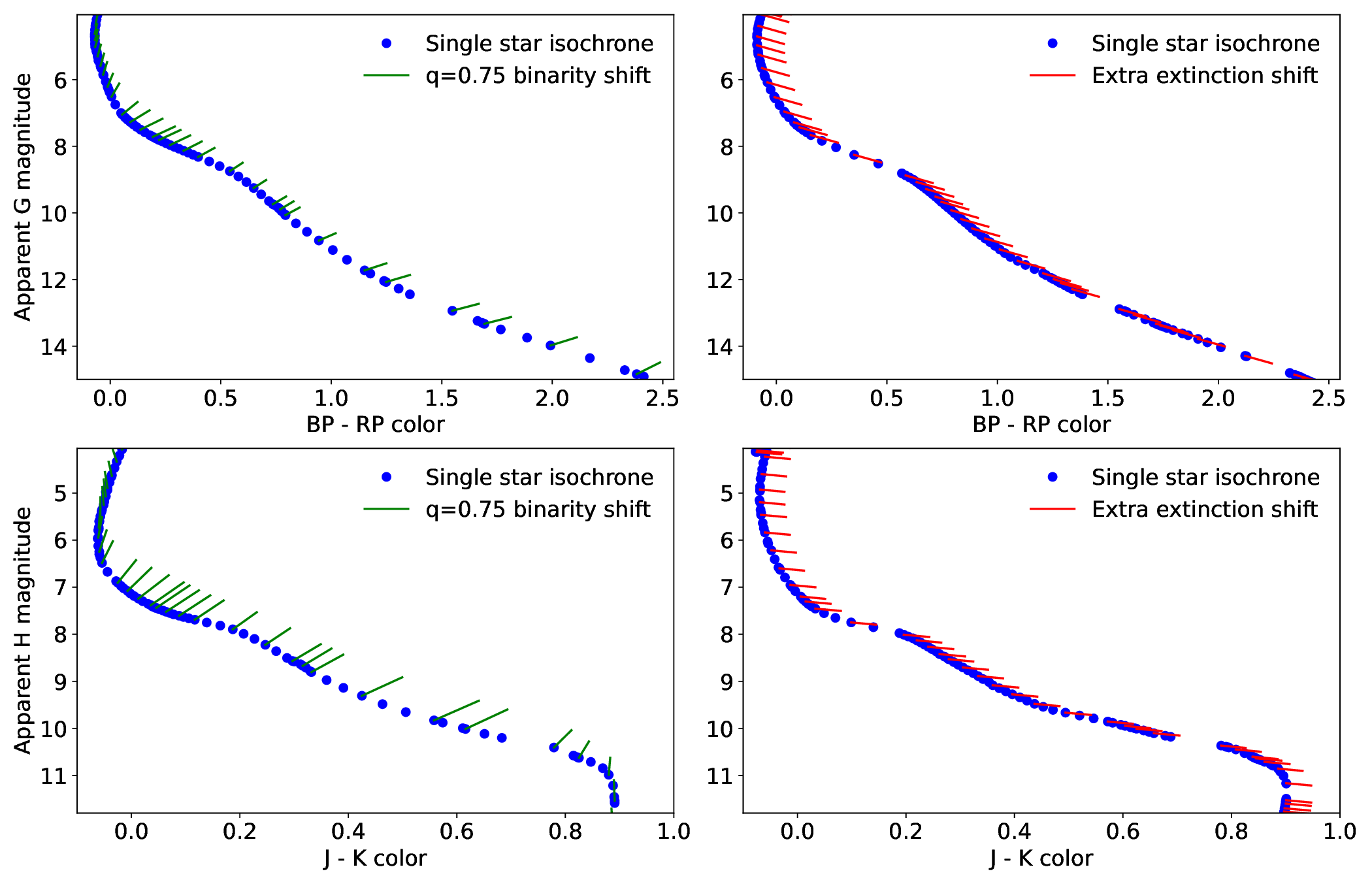}
\caption{Left: shift of binary systems with $q=0.75$ in the CMD relative to the single star sequence of \cite{2025AJ....169..116L}. For twin binaries ($q=1$), the color index is unchanged and the combined magnitude becomes brighter by 0.75 mag. Right: impact of enlarged absorption ($A_V=0.5$ relative to $A_V=0.23$ mag) on PARSEC 2.0 isochrones with star-by-star derived extinction \citep{1989ApJ...345..245C, 1994ApJ...422..158O}, calculated via the CMD 3.8 interface (\url{http://stev.oapd.inaf.it/cmd}), $R_V=3.1$.}
\label{fig:q075}
\end{figure*}

Importantly, binaries are not wiped out from this subsample completely, as some low-period ($P \lesssim 30$ days) systems show small RUWE values (Figure \ref{fig:SB}). The seven brightest stars with $G<5.4$ mag are missing from the subsample due to RUWE excess or a two-parameter solution in the case of Alcyone (25 Tau). Their absence is a disadvantage for isochrone fitting. Fortunately, the second-brightest Pleiades member, Atlas (27~Tau), is observed as a spectroscopic and interferometric binary, which allowed \cite{2025ApJ...990..107T} to derive a $136.2\pm1.4$ pc distance from the orbital solution alone. Hence, its absolute magnitude is reliable. Due to their weak dependence on color, high-mass binaries are almost indistinguishable from single stars in the CMD (Figure~\ref{fig:q075}), hindering detection of their multiplicity. 

\subsection{Binary stars in the CMD}
\label{q_calculation}

For coeval main sequence stars, an absolute magnitude directly relates to mass. Theoretical isochrones such as PARSEC \citep{2012MNRAS.427..127B, 2022A&A...665A.126N} or MIST \citep{2016ApJ...823..102C} are widely used to reproduce the observed color-magnitude relation and fit the cluster's age or multiplicity fraction. \cite{2025ApJ...979...92W} compared the observed sequences in Pleiades, Hyades, and Praesepe with theoretical  models and found similar deviation trends for these open clusters. The proposed $BP-RP$ color correction reaches its maximum of 0.25 and 0.15 mag for MIST and PARSEC models, respectively, in the $0.3-0.5~M_\odot$ range. \cite{2025AJ....169..116L} adjusted an empirical model using Gaia and 2MASS \citep{2006AJ....131.1163S} photometric data for Pleiades. It closely matches the subsample stars (Figure \ref{fig:CMD}) and was adopted for further calculations. 

The ultimate goal is to test if the observed photometric data are consistent with a single star, and if not, to estimate stellar masses and mass ratio. Table 3 from \cite{2025AJ....169..116L} provides an empirical mass -- absolute magnitude relation for single Pleiades stars. For unresolved binary systems, the total flux is considered to be a sum of primary and secondary component fluxes. Therefore, the combined magnitude is derived as:
\begin{equation}
\label{eq:combined_mags}
G=-2.5\log(10^{-0.4 G_1}+10^{-0.4 G_2})    
\end{equation}
where $G_1$ and $G_2$ are component magnitudes. Here it is assumed that both stars evolve as single ones depending on their masses $m_1$ and $m_2$. This approach may be incorrect for very close binaries, if the components have undergone mass transfer, which should be rare in a relatively young cluster. The synthetic 
magnitudes in Gaia and 2MASS passbands are calculated for a dense grid of masses in the $0.1 M_\odot<m_1<4.9 M_\odot$ range with a $0.001~M_\odot$ step. Mass ratios $q=m_2/m_1 \leq 1$ are sampled at 0.001 intervals. Their range depends on $m_1$, since the $m_2\geq0.1 M_\odot$ condition should be met. The observational sample includes $m_1\gtrsim 0.5M_\odot$ stars; therefore, substellar objects with $m_2\lesssim0.1 M_\odot$ 
can be neglected.

For sources with RUWE $<1.2$, astrometric data should be reliable and absolute magnitudes are derived from the reported Gaia DR3 parallaxes using Equation~\ref{eq:modulus}. The nominal uncertainties $\sigma_\varpi$ are used to calculate the inferred magnitude bounds assuming $\varpi\pm \sigma_\varpi$ parallax values. For other entries, a standard distance of 135.6~pc is adopted, which is a median for the low-RUWE subsample (Section~\ref{subsample}), while values of 133.7 and 138.5~pc constrain the error range. As shown by the example of Atlas with  RUWE $= 4.117$, the reported parallax can be misleading, and the cluster average better reflects its true distance. The $1/\varpi$ estimate  from $\varpi=8.12\pm0.48$~mas is around 123 pc; however, the orbital solution from \cite{2025ApJ...990..107T} securely puts Atlas inside the cluster's core with a $136.2\pm1.4$~pc value.

Differences between the observed ($G$, $BP$ and $RP$) and modeled ($\hat{G}$, $\hat{BP}$ and $\hat{RP}$) magnitudes are calculated with Equation \ref{eq:q}. The minimum photometric residual $\chi$ is selected to define a synthetic binary which has the closest match with the observed photometric data for sample objects. Similar calculations are made for the 2MASS photometric system in $J$, $H$ and $K$ bands.  

 \begin{equation}
  \label{eq:q}
     \chi =\sqrt{(G-\hat{G})^2+(BP-\hat{BP})^2+(RP-\hat{RP})^2}
      \end{equation}

\begin{figure*}[ht!]
\plotone{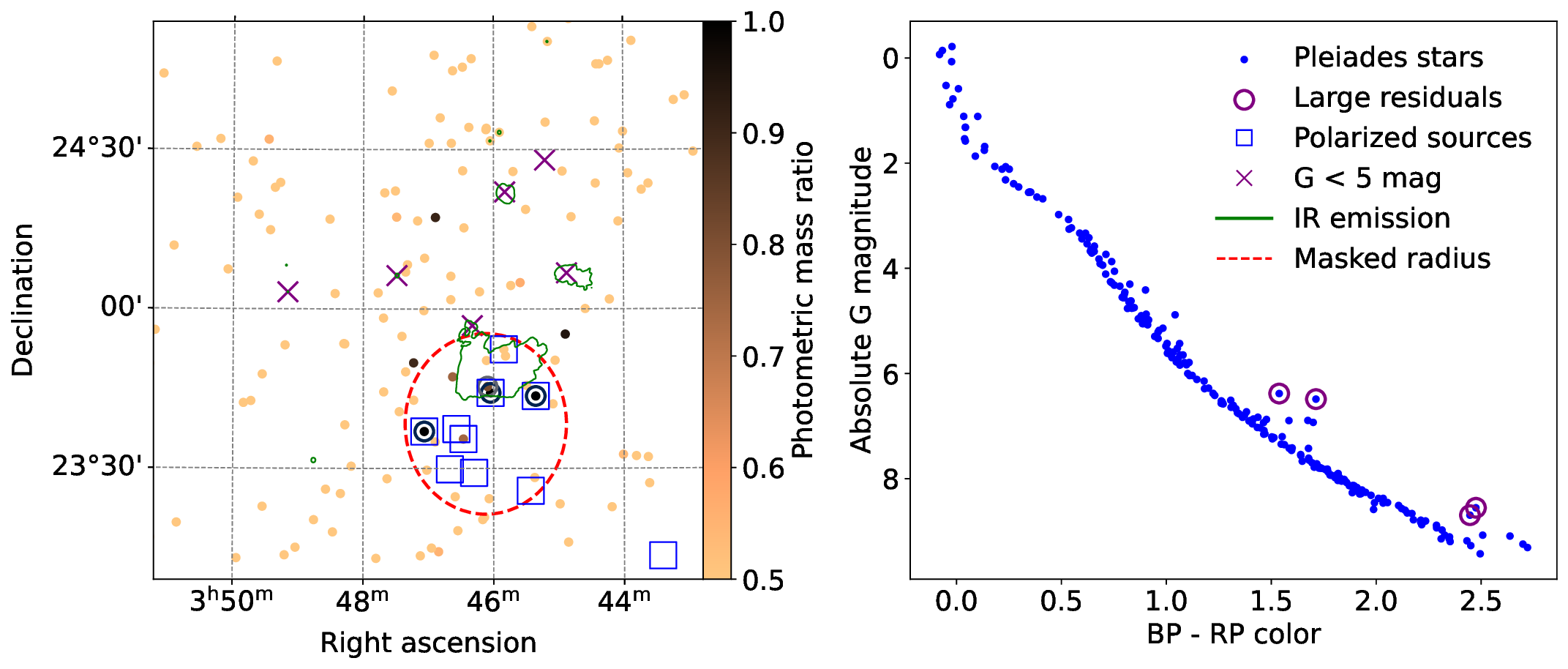}
\caption{Sky map of Pleiades stars with reliable astrometric and photometric data; the central part of the considered two degree radius is shown. Photometric mass ratio (Section \ref{q_calculation}) is color-coded, darker shades correspond to larger $q$. The four entries that are poorly fitted by either single-star or binary-star models (Section \ref{photometric_outliers}) are encircled; these are shown in the CMD (right panel). The six brightest Pleiades members are marked with purple crosses for guidance. Stars with linear polarization above 1\% threshold according to \cite{1986ApJ...309..311B} are enclosed in blue boxes; this includes sources with large RUWE. Nebulosity observed in the WISE W4 passband (22 $\mu m$, \cite{2010AJ....140.1868W}) is recovered using the HiPS2FITS service and shown with a green curve. A red circle encompasses the area of unreliable photometric $q$ estimation due to enlarged extinction (Section~\ref{excluded}).}
\label{fig:map}
\end{figure*}

For sources near the empirical single-star isochrone, Equation \ref{eq:q} minimization favors solutions with $q<0.4$ as expected. These entries are considered indistinguishable from single stars based on Gaia photometry. Out of 231 isolated low-RUWE sources, a $q>0.4$ solution is preferred for 35 entries and $q>0.5$ for only 20 (8.7\%) of them. Of these, just five systems are known as spectroscopic binaries (Section \ref{SB}).  The value of photometric residual $\chi$ is within 0.006 mag for all but four of the entries with $q>0.4$; this is likely a consequence of increased extinction in a relatively small cluster area (Section \ref{photometric_outliers}, Figure~\ref{fig:map}). The median residual is almost eight times larger if 2MASS photometry is adopted instead of Gaia for the same group of 35 objects. The lower accuracy limits its relevance for binarity parameter estimation, and only Gaia photometry is further considered. 

\subsubsection{Resolved systems}
\label{resolved}
An important precaution concerns binaries in the $0.1\lesssim \rho \lesssim\ 2$ arcsec separation range as they are effectively in the semiresolved regime in Gaia DR3. Depending on separation, their reported $G$ magnitudes can be related to a sole star, while $BP$ and $RP$ fluxes are combined from both components. Moreover, $BP$ and $RP$ fluxes can be affected at larger distances in the vicinity of a bright neighbor. Consequently, such systems appear as outliers in color-color diagram \citep{2024AJ....168..156C}, and the straightforward use of Equation~\ref{eq:q} brings an incorrect result. Mass ratios of the resolved systems are estimated in \cite{2025AJ....169..145C}. For nine sample entries, $BP$ and $RP$ magnitudes are not available due to interference from another Gaia source within $4\arcsec$.

\subsection{Validation for SB2 systems}
\label{photometric_validation}

\begin{figure*}[ht!]
\plotone{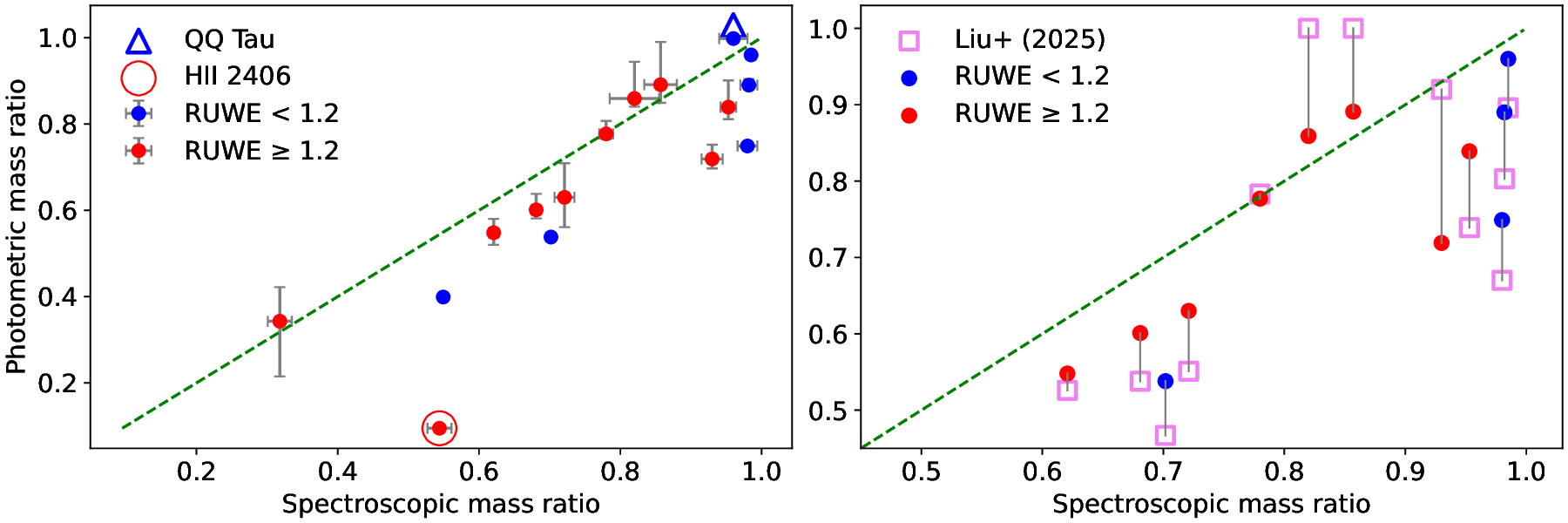}
\caption{The mass ratios of SB2 binaries (Table \ref{tab:SB2}) derived from the orbital solutions and CMD analysis. The diagonal line indicates the agreement of estimates. The error bars of photometric mass ratio account only for distance uncertainty; values of $q\lesssim0.5$ have particularly low credibility. For QQ~Tau, an enlarged extinction assumption is required to explain the photometric data (Section \ref{QQ_Tau}). The right panel shows a comparison with the best-fit result from \cite{2025AJ....169..116L} photometric analysis.} 
\label{fig:q_compare}
\end{figure*}

Double-lined spectroscopic (SB2) binaries  with orbital solutions  listed in Table \ref{tab:SB2} serve for validation of photometrically estimated $q_{\rm ph}$ (Figure \ref{fig:q_compare}). Based on 16 entries, it is lower than the spectroscopic value by $\Delta q=q_{\rm sp}-q_{\rm ph}= 0.09\pm0.12$, in line with the $\Delta q = 0.1$ mean offset from \cite{2025AJ....169..116L} (figure~8 in the respective paper), who used a Bayesian framework to derive mass ratios for the Pleiades stars based on Gaia and 2MASS data. Despite the similar result, our samples are different, and among the entries appearing in both, only HD 23631B ($G=9.74$  mag) shows closer agreement in the \cite{2025AJ....169..116L} work. For this star, a brighter neighbor ($G=7.29$ mag) at $\rho=6.3\arcsec$ could slightly affect $BP$ and $RP$ reported fluxes. For another 11 entries, the results are similar or the estimate from this study has better agreement with spectroscopic~data~(Figure~\ref{fig:q_compare}).

An example of HII 2406 with $q_{\rm sp} \sim 0.54$ indicates the limitation of photometric binary inference. \cite{2021ApJ...921..117T} spectroscopically measured the flux ratio of $1.3\pm0.3\%$; however, its binarity is not recovered photometrically, including in the study by \cite{2025AJ....169..116L}. Sources with $q_{\rm sp}>0.6$ start to reveal themselves as binary stars, albeit with an underestimated $q_{\rm ph}$ value.  

A difference between spectroscopic and photometrically fitted $q$ persists in other clusters. \cite{2024ApJ...962...41C} obtained $\Delta q=q_{\rm sp}-q_{\rm ph}= 0.13$ mean underestimation and adopted $\Delta q = 0.2$ as effective uncertainty (figure 10 in the respective paper). The recovered mass ratio is also underestimated relative to a true one in simulations modeling the photometric behavior of unresolved binaries (figure 5 in \cite{2024MNRAS.527.8718W}). Notably sequences for binary systems with $q=0.9$ and $q=1$ almost coincide in the CMD leading to degeneracy for near-twin pairs (figure 5 in \cite{2024ApJ...971...71J}).

\subsection{Sources with large photometric residuals}
\label{photometric_outliers}
Four low-RUWE sources clearly appear above the single-star sequence (Figure \ref{fig:map}), yet their residual in Equation \ref{eq:q} is larger than 0.01 mag, indicating that binary star models poorly fit them. The outlier objects can be interpreted as triple stars. Indeed, \cite{2023AJ....165...45M} estimated no less than 10\% of non-single Pleiades stars are triple or quadruple. However, this explanation can be challenged by a combination of two factors: these sources have low RUWE and are located in a small area reported for excessive absorption.

Based on the dynamical stability criterion of triple systems \citep{2001MNRAS.321..398M}, the outer period should be larger than the inner one at least by a factor of 4.7 \citep{2021Univ....7..352T}.  As evident from empirical data (Figure \ref{fig:SB}) and simulations (Section \ref{forward}), binaries with $P\gtrsim 20$ days start to show RUWE excess. The inner orbit therefore should be very compact to meet the stability criterion and keep the~RUWE~low.

\subsubsection{QQ Tau}
\label{QQ_Tau}
The discussion on specific objects begins with QQ~Tau ($G=14.26$ mag), a known SB2 system with near-twin components and 2.46 days period \citep{2025A&A...698A...7F}. Its photometric residual, $\chi=0.03$ mag (Equation~\ref{eq:q}), is at least five times larger than for the vast majority of sources. Three possible explanations can be mentioned. One requires flux from a tertiary component. To meet dynamical stability constraints and simultaneously keep a low RUWE value (1.12), the outer period should be roughly within 12 -- 20 days, making this system a convenient target for follow-up spectroscopic observations. 

Another option is that isochrones designed for single stars  are inadequate for short-period binary components. There are three more SB2 systems (V1229~Tau, HII 761 and V888 Tau) with periods within 3.5 days (Table \ref{tab:SB2}), their photometrically fitted $q$ is underestimated compared to the spectroscopic $q$ following a general trend (Figure \ref{fig:q_compare}). Due to combination of low mass and period, QQ Tau has the smallest orbit size among~them.

The final and indeed the favored possibility is an extinction excess toward QQ~Tau affecting the photometric~$q$ estimation. This suggestion is justified by the observed linear polarization of  
QQ~Tau and the other stars in the vicinity (Figure \ref{fig:map}) discussed in Section~\ref{extinction}. 

\subsubsection{Other outlying sources}

The largest photometric residual (Equation \ref{eq:q}), ${\chi=0.13}$ mag, belongs to HII 870 ($G=12.13$~mag), which has a reliable astrometric solution with RUWE~${=1.095}$. Its proper motion, parallax, and radial velocity from Gaia and ground-based observations are consistent with cluster membership. \cite{1997A&A...323..139B} detected a faint companion at $\rho=0.5\arcsec$   with $\Delta J = 4.4$~mag and estimated $q=0.24$. Whether it is a physical or optical pair, this component's luminosity is negligible. Companions with a noticeable flux in the 880 nm passband are ruled out for $\rho>0.1\arcsec$ separations by speckle observations \citep{2025AJ....169..145C}. 

V811 Tau appears within $10\arcmin$ from the previous source, has exactly the same RUWE and a similar magnitude of $G=12.07$  mag, while its photometric residual, $\chi=0.03$~mag, is smaller. Unlike for HII 870, the large uncertainties of radial velocity measurements from LAMOST \citep{2025A&A...698A...7F} and SDSS \citep{2025AJ....170...96M} allow to suggest this star is not single.

2MASS J03460556+2345261 is a fainter star with $G=14.38$ mag located just $1.4\arcmin$  from HII 870. Its photometric residual of $\chi=0.01$ mag is moderate, however the shift in the CMD due to enlarged extinction is less prominent for faint objects at optical wavelengths (Figure \ref{fig:q075}). It has a large $J-K=1.13$ mag value (Figure \ref{fig:CMD}) which is a sign of large extinction rather than binarity.

HII 975 ($G=10.36$ mag) has a RUWE of 1.311 which is slightly above the threshold for reliable solution. This source has a large $\chi\sim 0.08-0.09$ mag residual regardless of whether the reported Gaia DR3 parallax or the cluster mean distance is adopted for absolute magnitude calculation. Along with QQ Tau, HII 870, and V811~Tau, HII~975 shows a polarization excess, discussed below.

\subsection{Extinction and polarization}
\label{extinction}
Interstellar extinction and reddening is a complex and often overlooked factor \citep{2024arXiv240101116M}. The extinction in a given passband depends both on the interstellar matter along the line of view and the source energy distribution, and affects stars of different temperatures differently. Due to proximity to the Sun and moderate galactic latitude ($b\sim -24^{\circ}$), Pleiades stars have a low extinction with $A_V\sim 0.1-0.15$ mag \citep{2015ApJ...804..146D}. The fiducial PARSEC model used by \cite{2025AJ....169..116L} adopts $A_{V}=0.135$~mag.

The extinction and reddening within the Pleiades is not uniform. Unrecognized differential extinction can be misinterpreted as the outcome of stellar multiplicity during CMD analysis \citep{2008AJ....136.1388T}. 
Several works attempting to derive reddening values for millions of stars are summarized in the SDSS 19  \textit{astraMWMLite} catalog \citep{2025arXiv250707093S}.
The results for the sample objects are contradictory and appear to have limited credibility for comparison between cluster members. For instance, for HD 23512, the reported values of color excess $E(B-V)$ are between 0.04 and 0.86~mag. 

Unlike extinction or reddening,  polarization is directly measured and can be used to trace the absorption. Linear polarization occurs when light experiences nonuniform dichroic absorption by aligned asymmetric dust particles \citep{2012JQSRT.113.2334V,2015ARA&A..53..501A}. It is loosely correlated with total extinction, the upper bound of the polarization fraction is around 3\% per one magnitude of $A_V$ \citep{1975ApJ...196..261S}. \cite{1986ApJ...309..311B} conveyed a large polarimetric survey of Pleiades and found a linear polarization fraction above 1\% for 10 sources, all located to the south of Merope star (Figure \ref{fig:map}). A larger than average extinction is expected for these objects. All stars with large photometric residual discussed in Section \ref{photometric_outliers} show a polarization excess, apart from 2MASS J03460556+2345261 with no data. 

HD 23512 ($G=8.04$ mag) is the only sample source with $G<10$ mag and a large polarization, it is considered a standard star \citep{1982ApJ...262..732H}. \cite{2024MNRAS.535.1586C} argued its polarization may be caused by a yet unconfirmed companion. Indeed, the large RUWE value of 4.28 leaves little doubt this star is not single. However, considering that neighboring sources also show large polarization, enlarged absorption by the interstellar medium seems a more plausible interpretation.
\cite{2007ApJ...663..320F} studied spectral energy distribution for this object from ultraviolet to infrared and derived color excess $E(B-V)=0.36\pm0.01$ mag with $R_V=3.2\pm0.2$, which suggests total extinction $A_V$ around 1 mag. However, an unaccounted companion could compromise this result. 

\subsection{Excluded region}
\label{excluded}

Stars with large observed polarization and, hence, extinction may have biased binary parameters estimates. Considering these sources are located in a relatively small region (Figure \ref{fig:map}),
it is convenient to separate them from other cluster members. A radius of $17 \arcmin$ around celestial coordinates $\alpha=56.53^{\circ}, \delta=23.64^{\circ}$ defines a problematic area. Diffuse WISE W4 22~$\mu m$ emission \citep{2010AJ....140.1868W} partially overlaps with this region. 

The designated area is outside of the cluster core and should not have peculiar multiplicity properties. It includes 26 Gaia DR3 sources; 17 (65\%) of them have no close neighbors and RUWE $<1.2$ (Section \ref{subsample}). The corresponding numbers for the remaining cluster part are 384 and 214, yielding a proportion of 56\%. Among low-RUWE isolated sources, the fraction of entries with large photometrically inferred mass ratio is drastically different within and outside of the discussed area. For six out of 17 entries (35\%) inside the radius, $q_{\rm ph}>0.6$ is favored. The corresponding fraction outside is six out of 214 (2.8\%). Such a difference indicates that extinction excess has contaminated $q_{\rm ph}$ estimates, and the affected 26 sources are excluded from further statistical analysis. 

\subsection{Binary stars candidates}
\label{unresolved}

 \begin{figure*}[ht!]
\plotone{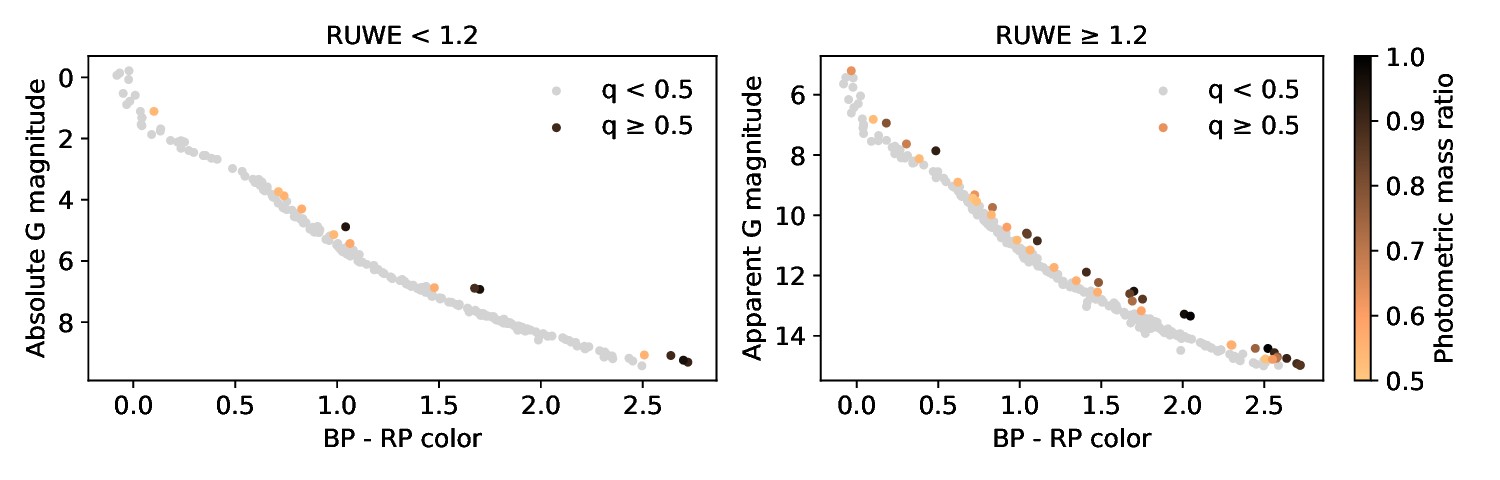}
\caption{Binary star candidates inferred from the CMD analysis; sources with resolved companions within $4\arcsec$ are excluded \citep{2025AJ....169..145C}. The entries with low and high RUWE are considered in the left and right panels, respectively.}  
 \label{fig:photometric}
\end{figure*}

The binary candidates inferred from the CMD analysis are shown in Figure~\ref{fig:photometric}.  A visible excess of binaries at the faint end of the sample is caused by the truncation bias. For single stars the $G=15$ mag threshold corresponds to $0.5~M_\odot$, and the less massive stars miss the sample. However, unresolved binaries with the primary masses up to $m_1 \sim 0.42 M_\odot$ may reach the limiting magnitude. 

For the most massive stars, the impact of binarity is hardly noticeable both in the visible and infrared colors (Figure \ref{fig:q075}) and the results for the six brightest members with $G<5$ mag are spurious, even though they are reasonably close to the spectroscopically derived value for Atlas (Table~\ref{tab:SB2}). The photometrically inferred binarity with $q\sim 0.63$ for Pleione (28 Tau, $G=5.20$ mag) can also be an artifact. This is a known SB1 system with $q<0.2$  \citep{2010A&A...516A..80N}. The existence of another companion was proposed, but numerous searches have not succeeded so far  \citep{2024ApJ...962...70K}. 

\begin{figure*}[ht!]
\plotone{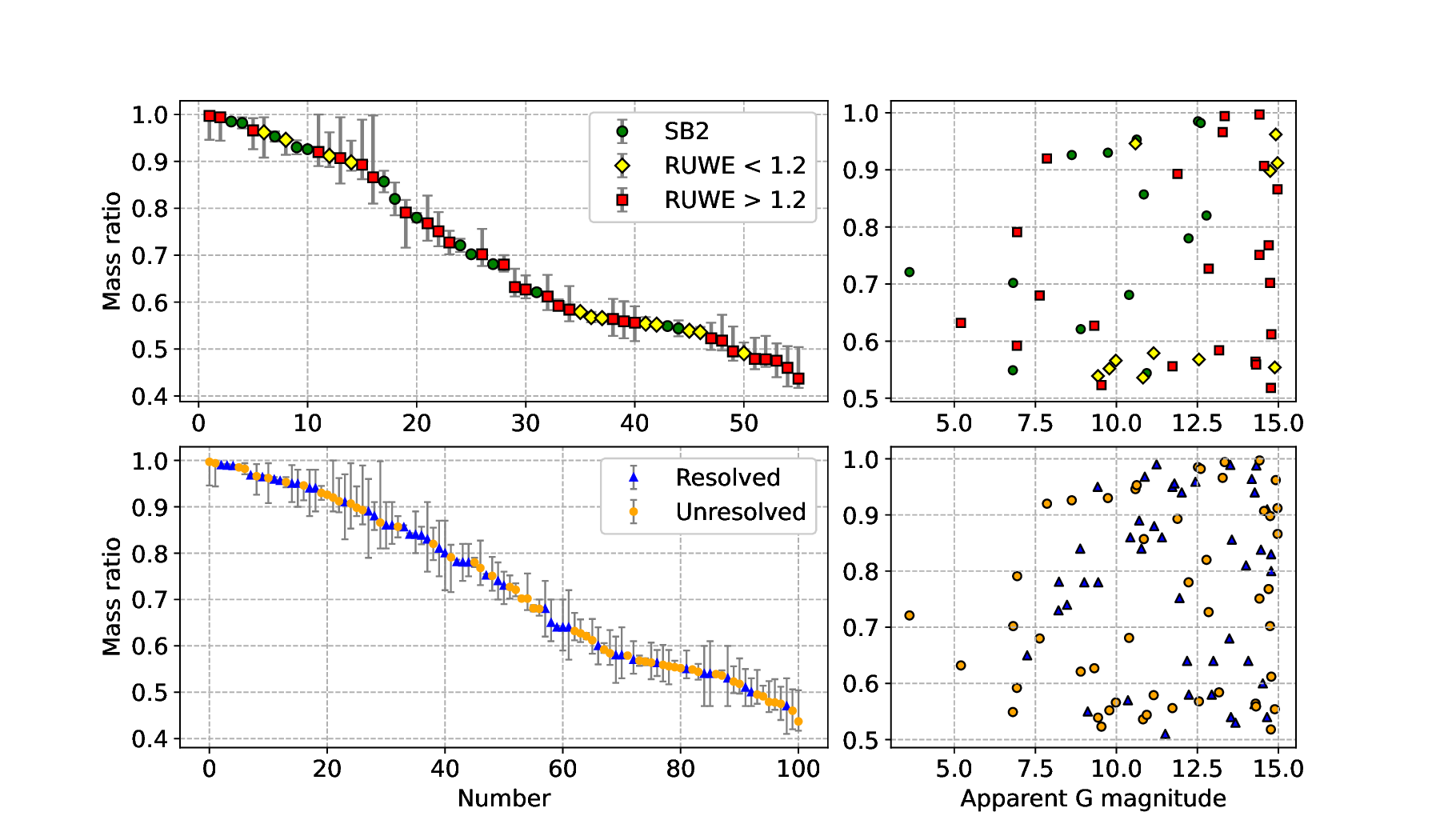}
\caption{The estimated mass ratios for unresolved binaries (top panel) and the whole binary population (bottom panel).}
\label{fig:all_binaries}
\end{figure*}

The inferred mass ratios for unresolved sources are plotted in Figure \ref{fig:all_binaries}. The errors shown account only for distance uncertainties, while the full uncertainties reach at least $\Delta q \sim 0.1$ (Section~\ref{photometric_validation}). For SB2 binaries with orbital solutions (Table~\ref{tab:SB2}), the spectroscopic~$q$ is more accurate and used instead of the photometrically derived value. The latter is particularly model-dependent for high-RUWE sources, and becomes incorrect if a given object is outside the 133.7 -- 138.5~pc  distance range. 

\section{Binarity census}
\label{census}
\subsection{Sample definition}
\label{sample_definition}
Census of binaries in the Pleiades  requires both resolved and unresolved systems to be counted. Pairs with $0.1 \lesssim \rho \lesssim 2$ arcsec separation are semiresolved in Gaia DR3, and standard procedures for photometric $q$ estimation should not be applied for them (Section~\ref{resolved}). The resolved systems are divided into two classes depending on whether Gaia DR3 has separate entries for both components. The mass ratio is estimated from the reported $G_1$ and $G_2$ magnitudes or flux contrast measured in speckle observations when $G_2$ is unavailable. 

At least 16 multiple systems with three or more components are identified in the sample. All of them include resolved companions; their configurations and parameters are discussed in the Appendix of \cite{2025AJ....169..145C}. For the binary fraction estimate, each system is counted only once. Hierarchical triple systems consist of an inner pair with stars Aa and Ab, and an outer component B (occasionally, star B is more massive than the inner pair components). Mass ratios of Aa -- Ab and Aa~--~B pairs are compared, and the larger one is selected for statistics. If both inner pair components have masses below the $0.5~M_\odot$  threshold, then the outer $q$ is considered. An unknown third component is possible in presumed  binary systems and may disturb $q$ estimation.

The input list of cluster members includes 410 Gaia DR3 entries (Section \ref{Sample}); 20 of them are deemed secondary components within Gaia-resolved  systems, reducing the number of unique systems to 390. After omitting objects from the area of high extinction (Section \ref{excluded}), 364 sources remain. Finally, 16 systems with $m_1<0.5 M_\odot$ are exempted from the binary fraction calculation to eliminate the 
truncation bias, leaving 348 entries, listed with their adopted mass ratios in Table~\ref{tab:main}.

\subsection{Binary fraction}
\label{fraction}

In the resolved multiplicity survey \citep{2025AJ....169..145C} the binary fraction $f = 6.7\%^{+1.8}_{-0}$ for $q>0.6$ systems or $f=9.8\%^{+0.3}_{-1.0}$ for $q>0.5$ within the 27~--~1350 au projected separation range was obtained. Under crude assumptions it was extrapolated to $f\sim 17\%$ for $q>0.6$ or $f\sim25\%$ for $q>0.5$ binaries across all orbit sizes. Following this study, more robust constraints can be placed. Summing up the resolved and unresolved systems, $57^{+9}_{-2}$ binaries with $q>0.6$ are counted, corresponding to $f=16.4\%^{+2.6}_{-0.6}$ fraction, or $81^{+7}_{-5}$ for $q>0.5$, yielding $f=23.3\%^{+2.0}_{-1.5}$ for cluster members with $m_1 > 0.5 M_\odot$. These align with the prior estimates.

Photometric $q$ can be underestimated up to $\Delta q \sim 0.1$ (Section \ref{photometric_validation}). The less severe bias exists for speckle observations (section~6.2 in \cite{2025AJ....169..145C}). No corrections were applied, so the estimated binary fraction may represent the lower bound of the real value.

\subsection{Comparison with other works}

\newcommand\Donada{{\cite{2023A&A...675A..89D}}}

\begin{deluxetable}{ccccc}
\tablecaption{Pleiades binary fraction in the literature}
\label{tab:comparison}
\tablehead{
    \colhead{ $q_{\rm min}$}     & \colhead{$m_1$} & \colhead{$f$} &\colhead{Ref}&\colhead{$f$, this study}}
    \startdata
     &$M_\odot$&\%&&\%
     \\
     --&0.11 -- 4.90 & $34\pm2$&L25& $>41$ ($m_1>0.5M_\odot$)\\
     --&$\sim$ 0.5 -- 5 & $>57$&T21&$>41$ \\
     --&0.57 -- 3.75 & $41\pm4$ & N20&$>40$\\
     --&0.57 -- 3.75 & $54\pm4$&L25\\
     -- &0.5 - 1.8 & $73\pm3$&M23&$>41$\\
     --&0.50 -- 0.90 & $45\pm5$&L25&$>37$\\ 
     --&0.90 -- 1.60 & $55\pm7$&L25&$>47$\\ 
    0.4&0.5 -- 1.2 & $19\pm1$ & P23&23 ($q>0.5$)\\
    0.5&0.28 -- 3.19 & 13 & J24&23 ($m_1> 0.5 M_\odot$)\\
    0.5&0.24 -- 1.66&$17\pm2$&A25&24 ($m_1> 0.5 M_\odot$)\\
    0.5&0.6 -- 1.0 & $23\pm6$ & P03&$23\pm3$\\
    0.5&0.6 -- 1.0 &$20\pm3$&N20\\
    0.5&0.6 -- 1.0&$22\pm4$&L25\\
    0.6& 0.1 -- 5& $22\pm4$&L23&16 ($m_1 > 0.5 M_\odot$)\\
    0.6&0.18 -- 2.30&$9\pm1$&D23&17 ($m_1> 0.5 M_\odot$)\\
    0.6&0.18 -- 2.30&$11\pm1$&L25\\
    0.6&0.24 -- 1.66&$13\pm2$&A25&17 ($m_1> 0.5 M_\odot$)\\
    0.6& 0.4 -- 3.6&$14\pm2$&J21&16 ($m_1 > 0.5 M_\odot$)\\
    0.6& 0.4 -- 3.6&$12\pm2$&L25\\
    0.6& 0.6 -- 0.9 & $10\pm3$&J21&$14\pm2$\\
    0.6& 0.9 -- 1.2 & $15\pm4$&J21&$15\pm1$\\
    \enddata
    \tablecomments{Selected binary fraction estimates for the Pleiades compared with the present study. The lower limit on $f$ is given if authors adopt no cutoff on $q$. References: A25 -- \cite{2025MNRAS.536..471A}, D23 -- \Donada, J21 -- \cite{2021AJ....162..264J}, J24 -- \cite{2024ApJ...971...71J}, L23 -- \cite{2023ApJS..268...30L}, L25 --  \cite{2025AJ....169..116L}, M23 -- \cite{2023AJ....165...45M}, N20 -- \cite{2020ApJ...903...93N}, P03 -- \cite{2003MNRAS.342.1241P}, P23 -- \cite{2023AJ....166..110P}, T21 -- \cite{2021ApJ...921..117T}.}
    \end{deluxetable}

Several estimates of binary fraction in the literature are listed in Table \ref{tab:comparison} with an emphasis on recent results relying on Gaia DR3 data. Most of them cover a large number of clusters, while this study focuses on the Pleiades, allowing its in-depth investigation starting from the member selection in \cite{2024AJ....168..156C}. The binary fraction in the 0.6 -- 1.0~$M_\odot$ mass range, $f=23.4\%_{-1.9}^{+2.5}$ for $q>0.5$, perfectly agrees with the earlier $f=23\pm6\%$ value of \cite{2003MNRAS.342.1241P}. 

Contrary to clusters, there are few recent binarity estimates for the field stars. The field population is complicated by diverse ages and has a significant contribution from systems with white dwarfs, which requires model-dependent bias corrections. For the F, G, and K-type stars, \cite{2013A&A...553A.124R} obtained $f=17\pm 3\%$  for $q>0.6$, which aligns with $f=15.3\%_{-0.5}^{+1.7}$ binary fraction among 0.6 -- 1.5 $M_\odot$ Pleiades stars. 

This study does not consider $G>15$ mag sources ($M \lesssim 0.5 M_\odot$), as their member selection is obstructed by the lack of radial velocity data. Since many authors include low-mass stars in the statistics, direct comparison is impeded. The literature estimates (Table \ref{tab:comparison}) are typically lower than in this study, possibly due to two factors. First, here the resolved binary population was separately considered, which makes an essential contribution for the nearby cluster. Another reason is a preferential loss of binary stars during membership classification due to less reliable astrometric solutions in Gaia. A remarkable exception is the $f=22\pm4\%$ estimate from \cite{2023ApJS..268...30L}. It appears to include entries with photometric $q>0.6$ along with spectroscopic binaries and high-RUWE sources, many of which are below this threshold, thus yielding a relatively large value.

\subsection{Mass dependence and mass ratio distribution}
\label{f_q}

\begin{figure*}[ht!]
\plotone{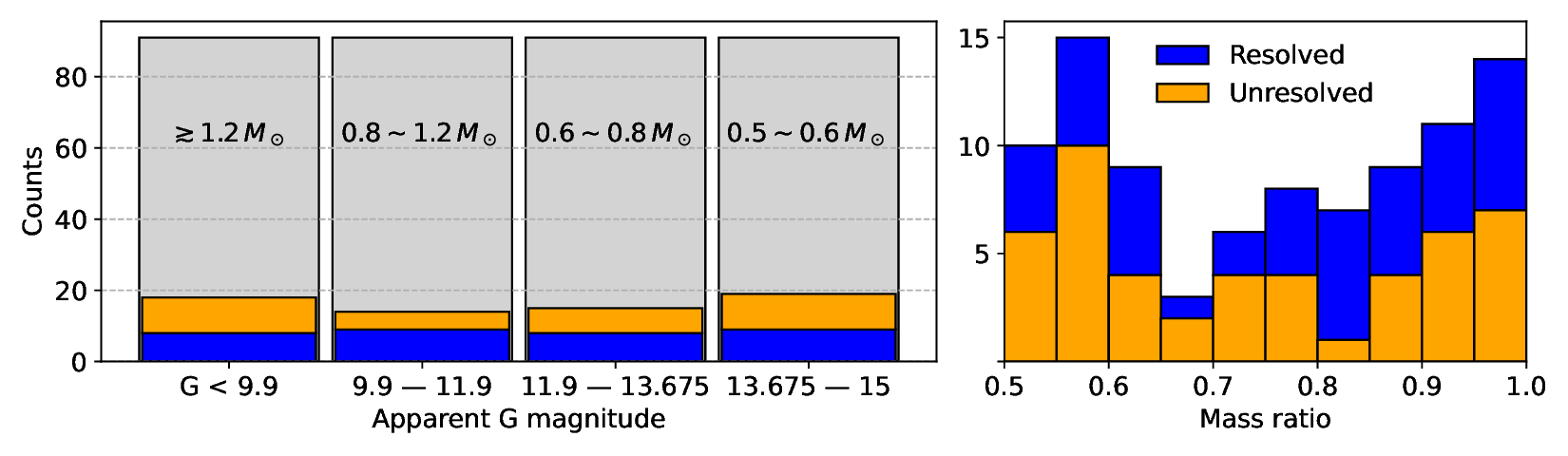}
\caption{Left: number of resolved and unresolved binaries with $q>0.6$ depending on source apparent magnitude. Approximate mass ranges for single stars are shown. Right: mass ratio distribution for binary stars with $q>0.5$ in the Pleiades.}  
\label{fig:histograms}
\end{figure*}

While the binary fraction of resolved systems was found to be distributed rather evenly over the magnitude bins, the unresolved sources produce a subtle minimum for solar-mass stars in Figure \ref{fig:histograms}. According to table~6 from \cite{2025AJ....169..116L}, the binary fraction in the Pleiades does not change with primary mass in the 0.5 -- 3 $M_\odot$ range for ${q>0.3}$ systems, set at $f\sim30\%$. In figure~9 of \cite{2021AJ....162..264J}, the binary fraction for $q>0.6$ binaries reaches a maximum of around 30\% at 1.5 $M_\odot$ and drops abruptly for less massive stars. In the field, the multiplicity fraction increases with stellar mass \citep{2023ASPC..534..275O}. However, \cite{2022A&A...657A..48S} concluded that it is more universal when a $q>0.6$ mass ratio cutoff is taken.

The mass ratio shows hints of a bimodal distribution with a dip around $q\sim 0.7$ in Figure \ref{fig:histograms}. It resembles the curve in figure 12 from \cite{2025AJ....169..116L} where the breakpoint was found at slightly larger $q=0.79\pm0.05$.  
A similar trend is seen for Pleiades and a few other clusters in figure 3 of \cite{2025MNRAS.536..471A}, where it was interpreted as a data reduction artifact. There is little agreement with figure~15 in \cite{2021ApJ...921..117T}, and the latter histogram is more consistent with a flat distribution. The obtained results contradict the decreasing function in figure 4 of \cite{2023AJ....165...45M}.
Star formation simulations from a giant molecular cloud by \cite{2024ApJ...977..203C} show in figure 5 that mass ratio distribution forms a subtle minimum around $q\sim 0.7$ relative to the primordial population during the process of cluster assembly. For comparison, the observed mass ratio distribution in the field is considered flat in the $0.2<q<0.95$ range with a peak for twin binaries \citep{2010ApJS..190....1R}.

\subsection{Low-luminous companions}
A mass ratio cutoff was applied previously to select systems with significant secondary flux. With some caveats, it is possible to extend the survey toward less luminous companions. Throughout this study, only main sequence stars were considered. Three white dwarfs are possibly related to Pleiades \citep{2022ApJ...926..132H}. The presence of binary systems with white dwarfs is proposed \citep{2024ApJ...976..102G}, yet none of the candidates is conclusively confirmed. If existent, a degenerate star was once the more massive component in the pair. 

Entries with photometric $q\leq0.5$ are considered indistinguishable from single stars (Figure \ref{fig:photometric}), and parameter RUWE (Section \ref{ruwe}) becomes the main binarity indicator in the low $q$ range. 
There are $40\pm5$~sources ($f=11.5\pm1.4\%$) with inferred $q\leq0.5$ and RUWE value above 1.4. If a less stringent RUWE $>1.25$ threshold is used, this number increases to $54^{+5}_{-6}$ ($f=15.5\%^{+1.5}_{-1.7}$). As discussed in Section \ref{forward}, RUWE is presumably sensitive enough to astrometric deviations from some brown dwarfs, but not giant planets at the Pleiades distance. 

Additionally, there are few confirmed binaries with low photometric $q$ and RUWE. These include two SB1 binaries, two speckle-resolved pairs with $\rho<1\arcsec$ and five wider Gaia-resolved pairs with RUWE of primary component below 1.25. Overall, the binary fraction in the $q\leq0.5$ range is no less than $14.9\pm1.5\%$ adopting RUWE $>1.4$ threshold or $18.1\%^{+1.4}_{-1.7}$ for RUWE $>1.25$. 

Summing up all suspected binaries, the lower limit on total binary fraction is 38.2\% or 41.4\% depending on the  RUWE threshold. The actual value should be higher since certain systems such as short-period binaries with low $q$ remain  undetected astrometrically and photometrically.  All triple and higher-order systems were counted only once, meaning no more than $\sim60\%$ of Pleiades stars with $m>0.5 M_\odot$  are single. For the field $0.75 <M < 1.25~M_\odot$ sample, \cite{2017ApJS..230...15M} obtained that $63\pm4\%$ of stars have no companions with $q>0.1$. However, this comparison is crude, since $q<0.1$ systems  may show RUWE excess in the Pleiades (Figure \ref{fig:Forward}) and thus be counted as binaries.

\section{Conclusions}
\label{summary}

The major conclusions are drawn as follows:

\begin{itemize}
\item Binaries with near-twin components ($q\sim 1$) are more likely to get astrometric solution with small deviations and low RUWE relative to systems with intermediate mass ratios of $q\sim 0.5$. The latter are disfavored during cluster member selection. The induced bias varies between clusters and disturbs comparison of multiplicity statistics (Section~\ref{q_bias}).
\item Photometrically estimated mass ratio is underestimated relative to spectroscopic (Section \ref{photometric_validation}).
\item Parallaxes inferred from non-single star solutions (Section~\ref{nss}) show worse convergence around the cluster mean compared to the single star solutions in Gaia DR3 for the same sources (Figure~\ref{fig:nss}).
\item An area of enlarged extinction (Figure \ref{fig:map}) is traced by observed polarization; it compromises the photometric study of multiplicity (Sections \ref{photometric_outliers} -- \ref{excluded}).
\item The inferred binary fraction for $q>0.6$ systems is $f=16.4\%^{+2.6}_{-0.6}$, which is compatible with the field population estimate (Section \ref{fraction}), but larger than most literature values for the Pleiades (Table \ref{tab:comparison}). 
\item The binary fraction presumably does not increase with stellar mass in the 0.5 -- 1.2 $M_\odot$ range (Section~\ref{f_q}). The mass ratio has a bimodal distribution with a minimum around $q\sim 0.7$ (Figure~\ref{fig:histograms}).
\end{itemize}

\begin{deluxetable*}{ccl}
\tablewidth{0pt}
\tablecaption{Pleiades sample source list}
\label{tab:main}
\tablehead{
\colhead{Column} & \colhead{Units} & \colhead{Description} 
}
\startdata
 Gaia &  & Gaia DR3 designation \\
 Gmag & mag & Gaia DR3 G magnitude\\
 RUWE &  & Gaia DR3 parameter RUWE \\
 $\varpi$ &mas& Gaia DR3 reported parallax \\
 $q$&&Adopted mass ratio of the binary system\\
 $q_{\rm min}$&&Lower bound of $q$\\
 $q_{\rm max}$&&Upper bound of $q$\\
 Type&&Type of binarity: \\
 &&G: Resolved in Gaia DR3, I: Resolved with speckle observations \citep{2025AJ....169..145C},\\
 && S: SB2 (Table \ref{tab:SB2}), p: photometric, RUWE $<1.2$, P: photometric, RUWE $\geq1.2$ (Section \ref{cmd})\\
 $q^\varpi$&&Photometrically fitted $q$ (Section \ref{q_calculation}); 1/$\varpi$ distance is assumed\\
 $m_1^\varpi$&$M_\odot$&Best-fit mass of the primary star;  1/$\varpi$ distance is assumed\\
 $\chi^\varpi$&mag&Photometric residual (Equation \ref{eq:q}); 1/$\varpi$ distance is assumed\\
  $q^C$&&Photometrically fitted $q$ (Section \ref{q_calculation}); 135.6 pc distance is assumed\\
  $m_1^C$&$M_\odot$&Best-fit mass of the primary star; 135.6 pc distance is assumed\\
   $\chi^C$&mag&Photometric residual (Equation \ref{eq:q}); 135.6 pc distance is assumed\\
 Trun&&Primary mass truncation flag, $m_1<0.5 M_\odot$ (Section \ref{sample_definition}) [1,0] \\
 Ext&&High extinction area flag (Section \ref{excluded}) [1,0] \\
 Sec&&Secondary source flag (Section \ref{sample_definition}) [1,0]\\
 Mult&&Triple or higher-order system flag (Section \ref{sample_definition}) [1,0]\\
    nss&&Gaia DR3 non-single star solution flag (Table \ref{tab:nss}) [1,0]\\
 SB&&Spectroscopic binarity type, based on Table \ref{tab:SB2} and \cite{2021ApJ...921..117T} [3,2,1,0]   \\
 Wide&&Resolved companion flag \citep{2025AJ....169..145C} [1,0]\\
 BF&&Subsample adopted for binary fraction calculation (Section \ref{sample_definition}) [1,0]\\
\enddata
\tablecomments{The input list is based on table 2 of \cite{2024AJ....168..156C}. Stellar binarity is probed with several indicators; sources with an upper bound on adopted mass ratio $q_{\rm max}>0.5$ are covered. Photometric mass ratios (Section~\ref{cmd}) are considered unreliable and are omitted for bright stars ($G<5$~mag) and resolved systems (Section \ref{resolved}). Two options are explored for the adopted distance,  either based on reported parallax in Gaia DR3 or on the cluster mean value (Section \ref{q_calculation}). The inferred values of $q^\varpi$ or $q^C$ below 0.4 are set to zero as these sources are photometrically indistinguishable from single stars. Estimates of photometric mass ratio  for the high extinction area are deemed unreliable; these sources are exempted from the binary fraction calculation (Section~\ref{excluded}).\\
(This entire table is available under the following link: \url{https://github.com/chulkovd/Pleiades_revisited}.)}
\end{deluxetable*}

\begin{acknowledgments}
 This work was largely inspired by the collaboration with Boris Safonov and Ivan Strakhov over the speckle interferometric survey of Pleiades at the SAI MSU observatory. 
The author is grateful to Alexey Sytov and Gabriel Perren for the discussion on CMD analysis and Zephyr Penoyre for insights on forward RUWE modeling. The author thanks the referee for a fast and smooth review.

\end{acknowledgments}

%

\software{Aladin Lite \citep{2014ASPC..485..277B, 2022ASPC..532....7B}, Astrophysics Data System\footnote{https://ui.adsabs.harvard.edu/}}, astropy \citep{2013A&A...558A..33A,2018AJ....156..123A,2022ApJ...935..167A}, CDS Cross-Match Service \citep{2012ASPC..461..291B, 2020ASPC..522..125P}, 
GaiaUnlimited \citep{2024A&A...688A...1C}, hips2fits\footnote{https://alasky.cds.unistra.fr/hips-image-services/hips2fits}, ipyaladin \citep{2020ASPC..522..117B}, SIMBAD \citep{2000A&AS..143....9W}, TOPCAT \citep{2005ASPC..347...29T}, VizieR \citep{2000A&AS..143...23O}



\bibliography{sample7}{}
\bibliographystyle{aasjournal}



\end{document}